\documentclass{JHEP3} 

\JHEPspecialurl{http://jhep.sissa.it/JOURNAL/JHEP3.tar.gz}

\usepackage{multicol,bbm}

\newcommand\fverb{\setbox\fverbbox=\hbox\bgroup\verb}
\newcommand\fverbdo{\egroup\medskip\noindent%
            \fbox{\unhbox\fverbbox}\ }
\newcommand\fverbit{\egroup\item[\fbox{\unhbox\fverbbox}]}
\newbox\fverbbox

\newcommand{\dsf}{\displaystyle\frac}

\newcommand{\be}{\begin{equation}}
\newcommand{\ee}{\end{equation}}
\newcommand{\pz}{\partial_z}

\newcommand{\ypp}{f^{\prime\prime}}
\newcommand{\yp}{f^{\prime}}
\newcommand{\Zpp}{Z^{\prime\prime}}

\newcommand{\dzp}{\Delta^+}
\newcommand{\dzm}{\Delta^-}

\newcommand{\dzpm}{\Delta^\pm}

\newcommand{\mn}{p}

\newcommand{\tM}{\tilde m_q}
\newcommand{\ts}{\tilde\sigma}
\newcommand{\tp}{\tilde\mn}

\title{Mesons and nucleons from holographic QCD in a unified approach}

\author{Hyun-Chul Kim
        \\
Department of
Physics, Inha University, Incheon 402-751, Korea\\
    E-mail: \email{hchkim@inha.ac.kr}}
\author{Youngman Kim
        \\
Asia Pasific Center
for Theoretical Physics and Department of Physics, Pohang University
of Science and Technology, Pohang, Gyeongbuk 790-784, Korea
\\  E-mail: \email{ykim@apctp.org}}
\author{Ulugbek Yakhshiev
        \\
Department of Physics, Inha University,
             Incheon 402-751, Korea\\
Department of Nuclear and Theoretical Physics, National
University of Uzbekistan, Tashkent-174, Uzbekistan\\
    E-mail: \email{u.yakhshiev@nuuz.uzsci.net}}

\received{\today}       

\abstract{ We investigate masses and coupling constants of mesons
and nucleons within a hard wall model of holographic QCD in a
unified approach.  We first examine an appropriate form of fermionic
solutions by restricting the mass coupling for the five dimensional
bulk fermions and bosons. We then derive approximated {\it analytic}
solutions for the nucleons and the corresponding masses in a small
mass coupling region.  In order to treat meson and nucleon
properties on the same footing, we introduce the same infrared (IR)
cut in such a way that the meson-nucleon coupling constants, i.e.,
$g_{\pi NN}$ and $g_{\rho
  NN}$ are uniquely determined.
  The first order approximation with
respect to a dimensionless expansion parameter, which is valid in the
small mass coupling region, explicitly shows difficulties to avoid
the IR scale problem of the hard wall model. We discuss possible ways
of circumventing these problems.}

\keywords{AdS/QCD, hadrons}

\preprint{INHA-NTG-06/2009}

\begin{document}

\section{Introduction}
The AdS/CFT
correspondence~\cite{Maldacena:1997re,Gubser:1998bc,Witten:1998qj}
that relates a strongly coupled large $N_c$ gauge theory to a weakly
coupled supergravity provides a novel approach to understand
nonperturbative features of quantum chromodynamics (QCD) such as the
quark confinement and spontaneous breakdown of chiral symmetry
(SB$\chi$S).  Though no rigorous proof exists for such a
correspondence in real QCD, this remarkable idea has triggered a
great amount of theoretical works on possible mappings from
nonperturbative QCD to 5 dimensional (5D) gravity, i.e. holographic
dual of QCD.  There are in general two different ways of modeling
holographic dual of QCD (see, for example, a recent
review~\cite{Erdmenger:2007cm}): One way is to construct 10
dimensional (10D) models based on string theory of D3/D7, D4/D6 or
D4/D8 branes~\cite{Karch:2002sh,Kruczenski:2003be,
  Kruczenski:2003uq,Sakai:2004cn,Sakai:2005yt}.
   The other way is so-called a
\textit{bottom-up} approach to a holographic model of QCD, also
known as AdS (Anti-de Sitter
Space)/QCD~\cite{Erlich:2005qh,DaRold:2005zs,DaRold:2005vr} in which
a 5D holographic dual is constructed from QCD based on the general
wisdom of AdS/CFT, the 5D gauge coupling being identified by
matching the two-point vector correlation functions.  Despite the
fact that this bottom-up approach is somewhat on an \textit{ad hoc}
basis, it reflects some of most important features of gauge/gravity
dual. Moreover, it is rather successful in describing properties of
mesons (see, for example, a recent review~\cite{Erdmenger:2007cm}).

On the other hand, QCD is not a conformal theory, in particular, in
the low-energy region, so one should also incorporate this property
in constructing an effective AdS/QCD model.  Consequently, different
models have been developed in this bottom-up approach. In
refs.~\cite{Erlich:2005qh,DaRold:2005zs,DaRold:2005vr}, the size of
the extra dimension (also known as the compactification scale) was
fixed at the point that corresponds approximately to the QCD scale
parameter $\Lambda_{\rm QCD}$, i.e., an infrared (IR) cutoff
parameter was explicitly introduced.  It is usually interpreted as
the confinement scale that also breaks sharply the conformal
invariance. These AdS/QCD models are called the \textit{hard-wall
model}.  On the contrary, there is an alternative approach called a
\textit{soft-wall model} in which the conformal invariance is broken
smoothly by introducing the dilaton background field in the 5D AdS
space~\cite{Batell:2008zm,Batell:2008me,Andreev:2006vy}.

While both approaches describe meson properties such as masses,
Regge trajectories, and so on, one serious problem arises,
when it comes to fermions in the AdS/QCD.  In order to describe the
fermions, another IR cutoff was introduced in hard-wall
models~\cite{Contino:2004vy,Gherghetta:2000qt,Mueck:1998iz,
Hong:2006ta,Arutyunov:1998ve,Henningson:1998cd}. As shown in
ref.~\cite{Hong:2006ta}, one has to introduce the IR cutoff for the
nucleon with a different value from the mesonic case so that one may
reproduce the excited nucleon spectra. In fact,
ref.~\cite{Hong:2006ta} used quite a small number for that.  However,
when one calculates the meson-nucleon coupling constants, an
inconsistency arises~\cite{Maru:2009ux}.  In order to determine the
coupling constants consistently, one must use the \textit{same} IR
cutoff.  Otherwise, one cannot fully consider whole information on the
meson and nucleon wavefunctions.  Thus, in the present work, we want
to investigate the meson and baryon sectors on an equal footing with
the same IR cutoff taken into account. To this end, we consider an
  anomalous dimension  of the baryon operator. We will also rederive the
nucleon wavefunctions analytically in such a way that the analysis on
the IR cutoff becomes easier.

The present work is organized as follows: In section 2, we briefly
review a hard-wall model for mesons and nucleons. In section 3, we
derive analytically the 5D energy eigenvalues and wavefunctions for
the nucleons.  In section 4, we discuss the meson-nucleon coupling
constants.  The results are presented and discussed in section 5.
The last section is devoted to summary and conclusion.
\section{Hard-wall model with holographic mesons and nucleons}
We first briefly review a hard-wall model for mesons and for nucleons,
developed in refs.~\cite{Erlich:2005qh,DaRold:2005zs} and in
ref.~\cite{Hong:2006ta}, respectively.  The model has a geometry of 5D
AdS
\begin{equation}
ds^2 \;=\; g_{MN} dx^M dx^N \;=\; \frac1{z^2}(\eta_{\mu\nu}dx^\mu
dx^\nu-dz^2)\,,
\end{equation}
where $\eta_{\mu\nu}$ stands for the 4D Minkowski metric:
$\eta_{\mu\nu} = \mathrm{diag}(1,-1,-1,-1)$. The 5D AdS space is
compactified by two different boundary conditions, i.e. the IR
boundary at $z=z_m$ and the UV one at $z=\epsilon\to 0$.  Thus, the
model is defined in the range: $\epsilon\le z \le z_m$. Considering
the global chiral symmetry $\mathrm{SU(2)}_L\times\mathrm{SU(2)}_R$ of
QCD, we need to introduce 5D local gauge fields $A_L$ and $A_R$ of
which the values at $z=0$ play a role of external sources for
$\mathrm{SU(2)}_L$ and $\mathrm{SU(2)}_R$ currents respectively.
Since chiral symmetry is known to be broken to $\mathrm{SU(2)}_V$
spontaneously as well as explicitly, we introduce a bi-fundamental
field $X$ with respect to the local gauge symmetry
$\mathrm{SU(2)}_L\times\mathrm{SU(2)}_R$, in order to realize the
spontaneous and explicit breakings of chiral symmetry in the AdS
side. The current quark mass term $\bar{q}_L\hat{m}q_R$ with
$\hat{m}=\mathrm{diag}(m_u,\,m_d)$ breaks explicitly chiral symmetry,
while its spontaneous breakdown is understood by the finite vacuum
expectation value (VEV) of the quark condensate $\langle
\bar{q} q \rangle$ that is regarded as an
order parameter.  Thus, considering these two, we can construct
the bi-fundamental 5D bulk scalar field $X$ in terms of the current
quark mass $m_q$ and the quark condensate $\sigma$
\begin{equation}
X_0(z) \;=\; \langle X \rangle \;=\; \frac12(m_q z + \sigma z^3)
\end{equation}
with isospin symmetry assumed.  The current quark mass $m_q$ is
defined as $m_q=(m_u+m_d)/2$.

The 5D gauge action in AdS space with the scalar bulk field and the
vector field  can be expressed as
\begin{equation}
S_M \;=\; \int dz\,\int d^4x  \sqrt{G}\, \mathrm{Tr} \left[ |DX|^2 +3|X|^2
-\frac{1}{2g_5^2}(F_L^2 + F_R^2) \right],
\label{eq:eff_meson}
\end{equation}
where $\sqrt{G}=1/z^5$, $DX \;=\; \partial X - i A_L X + iXA_R$, and
$F_{L,R}^{MN}=\partial^M A_{L,R}-\partial^N
A_{L,R}-i[A^M_{L,R},\,A^N_{L,R}] $.  The $g_5$ stands for the 5D gauge
coupling and is fixed by matching the 5D vector correlation function
to that from the operator product expansion (OPE):
$g_5^2=12\pi^2/N_c$.  The 5D mass of the bulk gauge field $A_{L,R}$ is
determined by the relation $m_5^2=(\Delta - p)(\Delta + p
-4)$~\cite{Gubser:1998bc,Witten:1998qj} where $\Delta$
denotes the canonical dimension of the corresponding operator with
spin $p$. The 5D mass of the bulk gauge field turns out to be
$m_5^2=0$, which is natural because of gauge symmetry.  Note that the
vector and axial-vector gauge fields are defined as
$V=(A_L+A_R)/\sqrt{2}$ and $A=(A_L-A_R)/\sqrt{2}$ that are coupled at
the boundary to the vector current $J_{V}^{a\mu}=\bar{q}\gamma^\mu t^a
q$ and the axial-vector current $J_{A}^{a\mu}=\bar{q}\gamma^\mu\gamma_5 t^a
q$ with $\mathrm{tr}(t^at^b)=\delta^{ab}$, respectively.
The effective action describes the mesonic
sector~\cite{Erlich:2005qh,DaRold:2005zs} completely apart from exotic
mesons~\cite{Kim:2008qh}.

Coming to the flavor-two ($N_F=2$) baryonic sector, one needs to
introduce a bulk Dirac field corresponding to the nucleon at the
boundary~\cite{Mueck:1998iz,Henningson:1998cd}. A specific hard-wall
model for the nucleon was developed by ref.~\cite{Hong:2006ta} and
was applied to describe the neutron electric dipole
moment~\cite{Hong:2007tf} and holographic nuclear
matter~\cite{Kim:2007xi}.  In this model, the nucleons are first
introduced as a massless chiral isospin doublets $(p_L,\, n_L)$ and
$(p_R,\,n_R)$ in such a way that the 't Hooft anomaly matching is
satisfied.  The spontaneous breakdown of chiral symmetry induces a
chirally symmetric mass term for nucleons
\begin{equation}
\mathcal{L}_{\chi SB} \;\sim\; -M_N \left( \begin{array}{c} \bar{p}_L \\
    \bar{n}_L  \end{array}\right)\, \Sigma\, (p_R, \,n_R) + \mathrm{h.c.},
\end{equation}
where $\Sigma=\exp(2i\pi^a\tau^a/f_\pi)$ is the nonlinear
pseudo-Goldstone boson field that transforms as $\Sigma\to U_L \Sigma
U_R^\dagger$ under $\mathrm{SU(2)}_L\times\mathrm{SU(2)}_R$.  The
$\tau^a$ and $f_\pi$ represent the SU(2) Pauli matrices and the pion
decay constant, respectively. Thus, we have to consider the following
mass term in the AdS side
\begin{equation}
\mathcal{L}_{\mathrm{I}} \;=\; -g \left( \begin{array}{c} \bar{p}_L \\
    \bar{n}_L  \end{array}\right)\, X\, (p_R, \,n_R) + \mathrm{h.c.},
\end{equation}
where $g$ denotes the mass coupling (or Yukawa coupling) between $X$
and nucleon fields, which is usually fitted by reproducing the nucleon
mass $M_N=940$ MeV. In this regard, we can introduce two 5D Dirac spinors
$N_1$ and $N_2$ of which the Kaluza-Klein (KK) modes should include the
excitations of the massless chiral nucleons $(p_L,\, n_L)$ and $(p_R,\,n_R)$,
respectively.  By this requirement, one can fix the IR boundary
conditions for $N_1$ and $N_2$ at $z=z_m$.

Note that the 5D spinors $N_{1,2}$ do not have chirality. However, one
can resolve this problem in such a way that the 4D chirality is
encoded in the sign of the 5D Dirac mass term. For a positive 5D mass,
only the right-handed component of the 5D spnior remains near the UV
boundary $z\to 0$, which plays the role of a source for the
left-handed chiral operator in 4 dimension. It is vice versa for a
negative 5D mass.  The 5D mass for the $(d+1)$ bulk dimensional spinor
is determined by the AdS/CFT expression
\begin{equation}
(m_5)^2 \;=\; \left(\Delta -\frac{d}{2}\right)^2.
\label{eq:5dmass}
\end{equation}
Since the nucleon consists of three valence quarks, the corresponding
$m_5$ turns out to be $m_5=5/2$. However, since QCD does not have
conformal symmetry in the low-energy regime, the 5D mass might acquire
an anomalous dimension due to a 5D renormalization flow. Though it is
not known how to derive it, we will introduce later some anomalous
dimention of the $m_5$ to see its effects on the spectrum of the
nucleon.

Considering all these facts, we are led to the 5D gauge action for the
nucleons
\begin{eqnarray}
S_N &=& \int dz\,\int d^4x  \sqrt{G}\, \mathrm{Tr} \left[{\cal
    L}_{\mathrm{K}}+{\cal L}_{\mathrm{I}} \right], \cr
{\cal L}_{\mathrm{K}} &=&  i \bar{N}_1
 \Gamma^M \nabla_M N_1 + i \bar{N}_2
 \Gamma^M \nabla_M N_2 - \frac52 \bar{N}_1 N_2 +  \frac52 \bar{N}_2 N_1
\nonumber\\
{\cal L}_{\mathrm{I}} &=& -g\left[ \bar{N}_1 X N_2 + \bar{N}_2
  X^\dag N_1\right],
\label{EffAct}
\end{eqnarray}
where
\begin{equation}
 \nabla_M \;=\; \partial_M + \frac{i}{4}\, \omega_M^{AB} \Gamma_{AB}
 -i A_M^L\,.
\end{equation}
The non-vanishing components of the spin connection are
$w_M^{5A}=-w_M^{A5}=\delta_M^A/z$ and
$\Gamma_{AB}=\frac1{2i}[\Gamma^A,\Gamma^B]$ are the Lorentz generators
for spinors.  The $\Gamma$ matrices in AdS are related to the 4D
$\gamma$ matrices as $\Gamma^M=  (\gamma^\mu,\,-i\gamma_5)$.
\section{Energy eigenvalues and eigenfunctions for nucleons}
Since the mesonic sector is well studied in
refs.~\cite{Erlich:2005qh,DaRold:2005zs}, we concentrate in this work
on the nucleon properties from the AdS/QCD model. In order to find the
mass spectrum of 4D nucleons, we have to solve the eigenvalue
equations that arise from expanding $N_1$ and $N_2$ in terms of the KK
eigenmodes.  Decomposing the $N_1$ and $N_2$ fields into the following
separable forms
\begin{eqnarray}
N_1(x,z) &=& f_{1L}(z) \psi_L(x) +  f_{1R}(z) \psi_R(x),\cr N_2(x,z)
&=& f_{2L}(z) \psi_L(x) +  f_{2R}(z) \psi_R(x),
\end{eqnarray}
where $\psi_{L,R}$ denote the components of the $4D$ eigen-spinors
$\psi=(\psi_L,\,\psi_R)^T$ with eigenvalues $p$, we obtain the
coupled equations for $f_{1L,1R}$ and $f_{1R,2R}$
\begin{eqnarray}
\left(\begin{array}{cc}
\pz-\frac{\dzp}{z}&-\phi(z)\\
-\phi(z)&\pz-\frac{\dzm}{z}\end{array}\right)\left(
\begin{array}{l}
f_{1L}\\
f_{2L}\end{array}\right)&=&-\mn\left(
\begin{array}{l}
f_{1R}\\
f_{2R}\end{array}\right)\,, \label{first}\\
\left(\begin{array}{cc}
\pz-\frac{\dzm}{z}&\phi(z)\\
\phi(z)&\pz-\frac{\dzp}{z}\end{array}\right)\left(
\begin{array}{l}
f_{1R}\\
f_{2R}\end{array}\right)&=&\mn\left(
\begin{array}{l}
f_{1L}\\
f_{2L}\end{array}\right)\,
\label{second}
\end{eqnarray}
with the IR boundary conditions $f_{1L}(z_m) \;=\; f_{2L}(z_m) \;=\;
0$, where $\Delta^{\pm}={2\pm m_5}$ and $\phi(z)=gX_0(z)z^{-1}$.

The left-handed eigenfunctions $f_{1L,2L}$ can be related to the
right-handed ones $f_{1R,2R}$ for given parity states. Introducing the
eigenvalues of the parity operator $P=\pm 1$, we can write the
relations~\cite{Hong:2006ta}
\begin{equation}
f_{2L}=-Pf_{1R}\,,\qquad f_{2R}=Pf_{1L}.
\label{eq:parity}
\end{equation}
Eq.(\ref{eq:parity}) being used, eqs.~(\ref{first}) and (\ref{second})
are reduced to
\begin{eqnarray}
\left(\pz-\frac{\dzp}z\right)f_{1L}&=&-(\mn+P\phi)f_{1R}\,,\label{EoneA}\\
\left(\pz-\frac{\dzm}{z}\right)f_{1R}&=&(\mn-P\phi)
f_{1L}\,.\label{EtwoA}
 \end{eqnarray}
These can be further decoupled as
\begin{equation}
\ypp+A(z)\yp+(\mn^2-B(z))f=0\,, \label{feq}
\end{equation}
where $f$ represents generically $f_{1L}$ and $f_{1R}$. The $f'$ and
$f''$ are the first- and second-order derivatives with respect to
$z$. The functions $A$ and $B$ are defined as
\begin{eqnarray}
A(z)&=&-\pz\ln\left[(\mn\pm P\phi)z^{\dzm+\dzp}\right]\,,\\
B(z)&=&\phi^2-\dsf{1}{z^2}\left(\dzm\dzp+\Delta_\pm\right)
\mp\frac{P\phi^\prime\dzpm}{z(\mn\pm P\phi)}\,.
\label{defS}
\end{eqnarray}
The function $f$ in eq.~(\ref{feq}) satisfies, respectively the UV and
the IR boundary conditions
\begin{equation}
f_L(\epsilon\equiv z_{\rm UV}\rightarrow 0)=0\,,\qquad f_R(z_m\equiv
z_{\rm IR})=0\,,\label{boundcond}
\end{equation}
which comes from the minimization of the
action~\cite{Contino:2004vy} and zeros of the right solutions $f_R$
will be used to generate the mass spectrum of the nucleons.

Since eq.(\ref{feq}) can be put into an equation of the
Sturm-Liouville type, we can write its solution in the form of
\footnote{We remark here that a similar method has been used in
\cite{KP2009}.}
\begin{equation}
f(z)=Z(z)\exp\left\{-\frac12\int^z A(z)dz\right\}=
Z(z)z^2\sqrt{\mn\pm P\phi}.
 \label{fform}
\end{equation}
Thus, we can consider eq.(\ref{feq}) as a simple quantum-mechanical 1D
potential-well problem. Furthermore, introducing the following
dimensionless variable and parameters
\begin{equation}
w=  zz_m^{-1}\,,\qquad  \tM =\frac{g}2\,m_qz_m\,, \qquad \ts
=\frac{g}{2}\,\sigma z_m^3\,,\qquad\tp=\mn z_m\,,
\end{equation}
we immediately obtain the dimensionless form of eq.(\ref{feq})
\begin{equation}
\Zpp(w)+\Big(\tp^2-U(w)\Big)Z(w)=0\,, \label{ME3}
\end{equation}
where the \textit{effective} potential is defined as
\begin{eqnarray}
U&=&U_0+U_1\,, \label{potb}\\
U_0&=& \dsf{(m_5\mp1)m_5}{w^2}-\frac{P\ts}{\tp}(2m_5\pm1) +\tM^2\,,\\
U_1&=&\pm\frac{\ts(\tM+\ts w^2)}{\tp(\tp\pm P(\tM+\ts
w^2))}\,(2m_5\pm1) \nonumber\\
&&+\frac{3\ts^2w^2}{(\tp\pm P(\tM+\ts
w^2))^2}+(2\tM+\ts w^2)\ts w^2\,. \label{pote}
\end{eqnarray}

Having examined the form of the solution given in eq.~(\ref{fform})
and the potential in eq.~(\ref{potb})-(\ref{pote}), we find that the
parameters $\tM$ and $\ts$ (and consequently $g$) are restricted by
the singularity of the potential $U_1$ and by the structure of
the corresponding even- or odd-parity solutions. Accordingly, the
input parameters are restricted as follows:
\begin{equation}
|\tM+\ts|=\dsf{|g|}{2}(m_q+\sigma z_m^2)z_m<\tp\qquad\mbox{or}
\qquad|g|<\frac{2\mn}{m_q+\sigma z_m^2}<g_{\rm crit}\,.
\label{parcon}
\end{equation}
Thus, we are able to restrict the mass coupling $g$ in the present
approach.  Note that if one restricts the value of $g$ by fitting it
to the mass of the lowest state $p_1 = N (940)$, which is given as
\begin{equation}
\frac{2\mn_1}{m_q+\sigma z_m^2}\equiv g_{\rm crit}, \label{gcrit}
\end{equation}
then the singularity in the potential for all other states can be
excluded. The meaning of the parameter-restriction condition
can be easily understood.  It measures the amount of the
corrections to the energy states when chiral symmetry is broken, as
obviously seen from eqs.~(\ref{EoneA})-(\ref{EtwoA}).

We now examine two different limiting cases in the mass coupling:
the limit of the \textit{small mass coupling} $|g| \approx 0$ and
the limit of the \textit{strong mass coupling} $|g|\approx g_{\rm
  crit}$. While in the limit of the small mass coupling the mass
spectrum of the nucleons are almost the same as those in the restored
phase of chiral symmetry, the changes in the spectrum turn out to be
rather large in the opposite limit.  A similar situation was already
studied in the Nambu-Jona-Lasinio model~\cite{Nambu:1961tp} (see also
relevant reviews~\cite{Klevansky:1992qe,Christov:1995vm}).

The normalizable zero-mode solutions in the chirally symmetric phase
is discussed in Ref.~\cite{Hong:2006ta}. After the spontaneous
breakdown of chiral symmetry, one still has zero modes for
fermions. To analyze this point, let us for the moment neglect the quark
mass ($m_q=0$) and look for the zero-mode solutions.  One can easily
find that the zero-mode equation has the form
\begin{equation}
\Zpp(w)-\left[\frac{n^2}{w^2}+\ts^2 w^4\right]Z(w)=0\,,\qquad
n^2\equiv (m_5\mp1)m_5\mp (2m_5\pm1)+3\,. \label{ZMeq}
\end{equation}
This equation can be solved analytically and has the solution
\begin{equation}
Z(w)=w^{1/2}\left[C_1 I_{m_5\mp 3/2}(\ts w)+C_2K_{m_5\mp 3/2}(\ts
w)\right]\,,
\end{equation}
where $I_{m_5\mp 3/2}$ and $K_{m_5\mp 3/2}$ represent the modified
Bessel functions. Taking into account eq.~(\ref{fform}), we obtain
the general zero-mode solution
\begin{equation}
f(w)=w^{5/2}\sqrt{\pm P\ts w^2}\left[C_1 I_{m_5\mp 3/2}(\ts
w)+C_2K_{m_5\mp 3/2}(\ts w)\right],
\end{equation}
which is proportional to the square root of the 5D mass
coupling. One can save or kill the left- or the right-handed zero
modes by appropriately choosing the sign of the mass coupling $g$.

If the conditions in eq.~(\ref{parcon}) are well satisfied, $U_1$
can be considered as a small perturbation of order
$O(\ts^2/\tp^2)$\footnote{See Eq.~(\ref{U1exp}). It is obvious that
$\tM/\tp \ll\ts/\tp$.}. To the first order, i.e., when $U_1=0$, the
problem can be solved analytically, so that
\begin{eqnarray}
Z(w)&=&w^{1/2}\left(CJ_{m_5\mp1/2}(a_{P}w)
+DY_{m_5\mp1/2}(a_{P}w)\right)\,,\label{firstorder}\\
a_P^2&=&\tp^2+\frac{P\ts}{\tp}(2m_5\pm1)-\tM^2\,,
\label{eq:nmass}
\end{eqnarray}
where $J_{m_5\mp1/2}$ and $Y_{m_5\mp1/2}$ are the Bessel functions
of the first and the second kinds, respectively. In order to get the
finite solutions at the UV boundary, the coefficients $D_{L,R}$ must
vanish\footnote{The value of the UV boundary is taken to be zero in
the present approach, i.e., $z_{\rm UV}=0$.}. The energy levels of
the states with a given parity correspond to the right-handed
solutions (see eq.~(\ref{boundcond})). Consequently, one is led to
the algebraic equations
\begin{equation}
\mu_n^2=a_P^2(\tp_n),
\label{Emain}
\end{equation}
where $\mu_n$ is the $n^{\rm th}$ zero of the Bessel function, i.e.,
$J_{m_5+1/2}(\mu_n)=0$.  Equation~(\ref{eq:nmass}) has a prominent
meaning. Since $\tilde{p}$ gives in general the whole spectrum of the
nucleon, one can immediately find that the mass of the first excited
state with positive parity turns out to be smaller than that with
negative parity. Thus, eq.(\ref{eq:nmass}) is consistent with the
experimental data. That is, the ordering of the nucleon spectrum is
analytically explained in this method.

The calculations of the next-order corrections are straightforward and
can be done by using the expansion parameters, $\tM/\tp$ and $\ts/\tp$,
and by expressing the potential $U_1$ in the form
\begin{eqnarray}
U_1^{(R)}&=&(2\tM+\ts w^2)\ts
w^2-\frac{\ts}{\tp^2}\left[(2m_5-1)(\tM+\ts w^2)-3\ts
w^2\right]\nonumber\\
&+&\frac{\ts}{\tp^2}\sum_{n=1}^\infty\left[3(n+1)\ts w^2-(\tM+\ts
w^2)(2m_5-1)\right]\left(\frac{\tM+\ts w^2}{\tp}\right)^n\,.
\label{U1exp}
\end{eqnarray}
It will be shown that the corrections are rather small, so that this
approximation works very well.

\section{Meson-baryon couplings}

Following Ref.~\cite{DaRold:2005zs}, one introduces the
gauge-fixing terms
\begin{eqnarray}
{\cal L}_{{\rm gf}}^V& =& - \frac{1}{2\xi_V g_5^2 z} \left[
\partial_\mu V^\mu -\xi_V h_1(V_5,z)
\right]^2,\\
{\cal L}_{{\rm gf}}^A &=& -\frac{1}{2\xi_A g_5^2z}
\left[
\partial_\mu A^\mu -\xi_A \big(h_1(A_5,z)+h_2(X,z)\big) \right]^2
\label{axialgf}
\end{eqnarray}
and can find an explicit form of the functions $h_{1,2}$ as done in
refs.~\cite{Hong:2006ta,Maru:2009ux}. In the unitary gauge
$\xi_{A,V}\rightarrow \infty$ the fifth component of the vector
field $V_5=(L_5+R_5)/2$ becomes infinitely heavy, so that it is
decoupled from the theory. Analogously, the linear combination of
the fifth component of axial-vector field $A_z=(L_5-R_5)/2$ and the
bi-fundamental scalar field $X=v\exp\{iP\}$ components $v$ and $P$
becomes infinitely massive. Introducing the corresponding relation
between $h_1(A_5,z)$ and $h_2(X,z)$ one can keep that linear
combination massless. As a result the massless pions can be
described. The corresponding equation for the pion mode function can
be analytically integrated in the chiral limit and has the form
\begin{equation}
f_\pi=\dsf{z^3}{N_0}\left[I_{2/3}(g_5\sigma
z^3/3)-\frac{I_{2/3}(g_5\sigma z_m^3/3)}{I_{-2/3}(g_5\sigma
z_m^3/3)}I_{-2/3}(g_5\sigma z^3/3)\right]\,,
\end{equation}
where $I_{\pm 2/3}$ are the modified Bessel functions. The
normalization condition for the pion mode function is fixed by a
canonical form of the kinetic term for pion fields
\begin{equation}
\int_0^{z_m} dz\left[\dsf{1}{2g_5^2z}f_\pi^2+\frac{z^3}{8X_0^2g_5^4}
\left(\partial_z\left(\frac{f_\pi}{z}\right)\right)^2\right]=1\,.
\end{equation}
Once the pion fields are correctly identified, then the $\pi NN$
coupling can be calculated as
\begin{eqnarray}
g_{\pi N^{(n)} N^{(n)}} &=& \int_0^{z_m} dz \frac{1}{z^4} \left[
f_{\pi}(f_{1L}^{(n)*}f_{1R}^{(n)}-f_{2L}^{(n)*}f_{2R}^{(n)})\right.\nonumber\\
&&\left.-\frac{gz^2}{2X_0g_5^2}\,\partial_z \left (
\frac{f_{\pi}}{z} \right
)(f_{1L}^{(n)*}f_{2R}^{(n)}-f_{2L}^{(n)*}f_{1R}^{(n)}) \right]\,.
\label{pinn}
\end{eqnarray}
Similarly, $\rho NN$ coupling have the form of~\cite{Maru:2009ux}
\begin{eqnarray}
g_{\rho N^{(n)} N^{(n)}} &=& \int_0^{z_m} dz \frac{1}{z^4} \left[
f_\rho + c z \partial_z f_\rho \right] \left[ |f_{1L}^{(n)}|^2 +
|f_{1R}^{(n)}|^2 \right]\,,
\end{eqnarray}
where $c$ is the constant of order in unity and the normalized
$\rho$ meson wave function has the following form
\begin{equation}
f_\rho=\frac{zJ_1(m_\rho z)}{\left(\int_0^{z_m}dzz[J_1(m_\rho
z)]^2\right)^{1/2}}\,.
\end{equation}
For completeness, we remind here the normalization condition for the
mode functions of the baryons
\begin{equation}
\int_0^{z_m}\dsf{dz}{z^4}\left(|f_{1L}^{(n)}|^2+|f_{2L}^{(n)}|^2\right)=1=
\int_0^{z_m}\dsf{dz}{z^4}\left(|f_{1R}^{(n)}|^2+|f_{2R}^{(n)}|^2\right)\,.
\end{equation}
\section{Results and discussions}
We now present the results of this work and discuss them.
Most of input parameters of the model such as $m_q$, $\sigma$ and
$z_m$ are quite well fitted in the mesonic
sector~\cite{Erlich:2005qh}. Hence, we have only one free parameter $g$
to reproduce the data in the baryonic sector. However, the IR cutoff
$z_m$ in the baryonic sector, which is often interpreted as a scale of
the confinement, takes different values from those in the mesonic
sector.  Actually, ref.~\cite{Hong:2006ta} performed two different
fittings of these parameters. In the first fitting of
ref.~\cite{Hong:2006ta}, the $z_m$ and the $\sigma$ were fixed in the
mesonic sector, and the $g$ is fitted to the nucleon mass.
In the second fitting, the $z_m$ and the $g$ were taken
respectively to be $(205\,\mathrm{MeV})^{-1}$ and $14.4$ such that
the masses of the nucleon and the Roper resonance $N(1440)$ were
reproduced.
Since there is no reason for a nucleon to have the same scale of the
confinement as that for a meson, this might be an acceptable argument
as far as one treats mesons and baryons separately. However, there is
one caveat. When it comes to some observables such as the meson-baryon
coupling constants, we need to treat the mesons and baryons on the
same footing and require inevitably a common $z_m$. Otherwise, we are
not able to consider whole information on both mesons and
baryons. Moreover, a model uncertainty brings on by the mass coupling
$g$. Thus, in the present section, we will carry
out the numerical analysis very carefully, keeping in mind all these
facts.

We first take different values of the $z_m$ from those in the
mesonic sector and try to fit the data as was done in
ref.~\cite{Hong:2006ta}.  In this case, $\sigma$ is defined as
$\sigma={4\sqrt2}({g_5z_m^3})^{-1}$. Furthermore, we will examine
two different limits of the mass coupling $g$: In the limit of the
small mass coupling, there are three free parameters $m_q$, $g$ and
$z_m$. All other parameters can be related to $z_m$. On the other
hand, in the limit of the strong mass coupling, the $g$ can be fixed
by eq.~(\ref{gcrit}), which leaves only two free parameters.
Obviously, the dependence on the current quark mass $m_q$ must be
tiny because of its smallness, so we can simply neglect it.  In this
case, we have only one free parameter.

\TABLE{
\begin{tabular}{cccccccccc}\hline\hline
$m_5$& $z_m^{-1}$ & $\sigma^{1/3}$ &$m_q$& $g$&
$(p,n)^+$ & $N^+(1440)$ &$N^-(1535)$&$\rho$(776)&$\rho$(1475)\\
\hline\hline
\multicolumn{10}{c}{The exact numerical results (the limit of the
  strong mass coupling)}\\
5/2& $130.9^*$& 126.4 & $0$ & 15.9& 940 &1336.2  &1366.5 &314.8&722.6\\
5/2&$129.7^*$ & 125.2 & $3^*$  & 15.7& 940 & 1328.2 & 1357.5&311.8&715.7\\
5/2&$126.0^*$ & 121.7 & $10^*$  & 15.4& 940 & 1304.6 & 1331.9&303.0&695.5\\
\hline
\multicolumn{10}{c}{The exact numerical results (the case of the small
  mass coupling)}\\
5/2& $147.2^*$& 142.1 & $0$ & 6.0$^*$& 940 &1440.3  &1456.6 &354.0&812.6\\
5/2&$147.0^*$ & 141.9 & $3^*$  & 6.0$^*$& 940 & 1439.3 & 1455.6&353.5&811.5\\
5/2&$146.3^*$ & 141.3 & $10^*$  & 6.0$^*$& 940 & 1434.6 & 1451.0&351.8&807.6\\
\hline
\multicolumn{10}{c}{The leading order approximation (the case of the
  small mass coupling)}\\
5/2&$147.2^*$ & 142.1 & $0$ &  6.0$^*$& 919 & 1428.4 & 1445.1 &354.0&812.6\\
5/2&$147.0^*$ & 141.9 & $3^*$  & 6.0$^*$& 918 & 1426.5 & 1443.1&353.5&811.5\\
5/2&$146.3^*$ & 141.3 & $10^*$  & 6.0$^*$& 914 & 1415.0 & 1436.6&351.8&807.6\\
\hline\hline
\end{tabular}
\caption{The results of the spectra of the nucleon and the $\rho$
meson. In the limit of the small mass coupling, there are three free
parameters $m_q$, $g$ and $z_m$, while in the limit of the strong
mass coupling, the $g$ is fixed near its critical value (see
Eq.~(\ref{gcrit})).  All dimensional quantities are expressed in
units of MeV. The asterisks indicate input parameters. The parameter
$\sigma$ is defined as $\sigma={4\sqrt2}({g_5z_m^3})^{-1}$. The
results of the leading-order approximation are yielded according to
Eq.~(\ref{Emain}). The 5D nucleon mass $m_5$ is given by the
AdS/CFT, eq.~\ref{eq:5dmass}. \label{tabH} } }
The results of the calculations are listed in table~\ref{tabH}.  In
the first part of the table, we present the results in the limit of
the strong mass coupling. They are more or less the same as those
obtained in ref.~\cite{Hong:2006ta}.  For comparison, we list the
results for the small mass coupling in the middle part of
table~~\ref{tabH}, and those of the leading-order approximation (see
Eq.~(\ref{Emain})) in the last part, respectively.
The mass coupling $g$ is chosen to be $6$ in both cases.
While the spectrum of the nucleon seems to be qualitatively well
reproduced, that of the $\rho$ meson is fairly underestimated in
comparison with the experimental data. In the case of the strong mass
coupling, the situation becomes even worse.  However, as dictated by
eq.~(\ref{eq:nmass}), the ordering of the nucleon-parity states are
correctly reproduced for $0<g<g_{\rm crit}$.

The results listed in table~\ref{tabH} indicate that it is not
possible to reproduce the spectra of the $\rho$ meson and the nucleon
at the same time.\footnote{Note that in the present work we do not aim
at the fine-tuning of the parameters to reproduce the experimental
data. The output data in baryonic sector is quite stable for changes
in $\sigma$.} As an attempt to improve the above-presented results,
we want to introduce an anomalous dimension of the 5D nucleon
mass. Note that the 5D mass of the bulk vector field does not
acquire any anomalous dimension because of the gauge symmetry.

Table~\ref{mesdata} lists the results of calculations for
different values of the 5D mass $m_5$, whereas the $z_m$, the
$\sigma$, and $m_q$ are fitted to the mesonic sector.
\TABLE{
\begin{tabular}{cccccccccc}\hline\hline
$m_5$&$z_m^{-1}$ & $\sigma^{1/3}$ &$m_q$& $g$&
$(p,n)^+$ & $N^+(1440)$ &$N^-(1535)$&$\rho$(776)&$\rho$(1475)\\
\hline\hline \multicolumn{10}{c}{Model A}\\
0&$323^*$ & $327^*$& 2.29$^*$&  -1.5 &1009 & 2029 & 2036&776&1783\\
1&$323^*$ & $327^*$& 2.29$^*$& 1 &1449 & 2495 & 2498&776&1783\\
2&$323^*$ & $327^*$& 2.29$^*$& 2.1 &1853 & 2936 & 2949&776&1783\\
5/2&$323^*$ & $327^*$& 2.29$^*$& 2.5 &2050 & 3150 & 3168&776&1783\\
\hline\hline \multicolumn{10}{c}{Model B}\\
0&$346^*$ & 308$^*$& 2.30$^*$& -2.4 &1081& 2031 & 2040&832&1910\\
1&$346^*$ & 308$^*$& 2.30$^*$&  1.3 &1553& 2494 & 2499&832&1910\\
2&$346^*$ & 308$^*$& 2.30$^*$&  3.1 &1985& 2937 & 2952&832&1910\\
5/2&$346^*$ & 308$^*$& 2.30$^*$&  3.6 &2196& 3151 & 3172&832&1910\\
\hline\hline
\end{tabular}
\caption{The results of the spectra of the nucleon and the $\rho$
meson with the 5D mass $m_5$ varied in the range of $0\le m_5\le
5/2$. The mass coupling $g$ is fitted to the spectrum of the nucleon.
All other parameters are taken from ref.~\cite{Erlich:2005qh}. All
dimensional quantities are presented in units of MeV. The asterisks
indicate input parameters.  Two different sets of input parameters are
used as in ref.~\cite{Erlich:2005qh}.
 \label{mesdata}}
}
Note that here the nucleon mass is not used as an input.  Varying the
value of $g$, we try to fit the spectrum of the nucleon.  We present
the results from two different parameter sets called model A and model
B. In this analysis, we take the values of the $z_m$ and $\sigma$ from
ref.~\cite{Erlich:2005qh}.  Note that the $\rho$ meson mass is used as
an input in model A, while model B corresponds to the global fitting
done in ref.~\cite{Erlich:2005qh}. The 5D nucleon mass is varied in
the range of $0\le m_5\le 5/2$, its anomalous dimension being
considered as mentioned before. As shown in table~\ref{mesdata}, the
best result is obtained with $m_5=0$. Though the absolute values of
the nucleons turn out to be overestimated in contrast to the previous
analysis presented in table~\ref{tabH}, the ground-state masses of the
nucleon and the $\rho$ meson are qualitatively well reproduced within
$30\,\%$.

We are now in a position to include meson-baryon coupling
constants in the present numerical analysis. We will consider here the
$\pi NN$ and the $\rho NN$ coupling constants in addition to the
$\rho$ meson and the nucleon spectra.  One has to keep in mind that in
order to calculate the meson-baryon coupling constants it is essential
to use the same $z_m$ for the mesonic and baryonic sectors. Otherwise,
it is not possible to keep whole information on the
wavefunctions. Thus, it is of utmost importance to compute all
observables with the same set of parameters. We perform a global
fitting procedure to obtain the results listed in
table~\ref{globfitG}.  Note that we consider here the chiral limit
($m_q=0$), since its effects on the results are rather tiny.
\footnote{Note that in the chiral limit, the nucleon mass is different
from experiments, $M_n\simeq 939~{\rm MeV}$. For instance,
$M_n\simeq 882~{\rm MeV}$ in the chiral limit~\cite{PMWHW}.}
\TABLE{
\begin{tabular}{cccccccccc}\hline\hline
$z_m^{-1}$ &$\sigma^{1/3}$  &$g$& $(p,n)^+$ & $N^+$
&$N^-$&$\rho$&$\rho$&$g_{\pi NN}$
&$g_{\rho NN}$\\
&&&(939)&(1440)&(1535)&(776)&(1475)&(13.1)&(2.4)\\
\hline\hline
$285^*$ & $256^*$&   -2.0$^*$ &890 & 1791 & 1797&685&1573&1.65&1.39\\
$285^*$ & $237^*$&    -2.0$^*$ &890 & 1790 & 1796&685&1573&1.76&1.39\\
$285^*$ & $256^*$&   -8.0$^*$ &930 & 1826 & 1856&685&1573&4.89&1.34\\
$285^*$ & $237^*$&   -8.0$^*$ &920 & 1817 & 1843&685&1573&5.12&1.35\\
\hline
$285^*$ & $227^*$&   -9.6$^*$ &930 & 1826 & 1856&685&1573&6.12&1.34\\
\hline
$280^*$ & $252^*$&  -2.0$^*$ &874 & 1760 & 1765&673&1546&1.65&1.39\\
$280^*$ & $233^*$&  -2.0$^*$ &874 & 1759 & 1764&673&1546&1.76&1.39\\
 \hline\hline
\end{tabular}
\caption{The results of the spectra of the nucleon and the $\rho$
meson and the $\pi NN$ and $\rho NN$ coupling constants.  The
parameters $z_m$, $\sigma$, and $g$ are found by the global fitting
procedure. The anomalous dimension of the
5D nucleon mass is chosen in such a way that the 5D mass vanishes.
All other definitions are the same as in tables~\ref{tabH}
and~\ref{mesdata}. \label{globfitG}}
}
We assume also that the 5D nucleon mass acquires a large anomalous
dimension so that it may vanish, i.e., $m_5=0$.  The best fit is
obtained with the parameters fitted as follows:
$z_m=(285\,\mathrm{MeV})^{-1}$, $\sigma=(227\,\mathrm{MeV})^3$, and
$g=-9.6$.  The masses of the ground-state nucleon and the $\rho$
meson are in good agreement with the experimental data. Moreover,
those of the excited states are qualitatively well reproduced within
$10-20\,\%$. However, the coupling constants are in general about
$50\,\%$ underestimated.  We mention that in ref.~\cite{Maru:2009ux}
the dependence of the meson-baryon coupling constants on the $z_m$
was investigated without considering hadron spectra but the results
for the coupling constants are more or less in the same level as in
the present work.
\section{Summary}
We have investigated the mesons and the nucleons in a unified
approach, based on a hard-wall model of
AdS/QCD~\cite{Erlich:2005qh,Hong:2006ta}. We first have decoupled
the equations of motion for the nucleons and casted them into the
Sturm-Liouville type such that the problem for the nucleons is
reduced to a simple one dimensional quantum-mechanical
potential-well problem. In order to study the nucleon spectrum, we
developed an approximated method in which the effective potential
can be expanded. The method of this approximation was shown to work
very well.  In particular, the correct ordering of the nucleon
parity states was {\it analytically} shown in this method.

We then have carried out various numerical analyses, varying the
model parameters such as the IR cutoff $z_m$, the quark condensate
$\sigma$, and the mass coupling (or Yukawa coupling) $g$. In order
to improve the results of the nucleon and the $\rho$ meson spectra
on an equal footing, we have introduced an anomalous dimension of
the 5D nucleon mass. We found that the zero 5D nucleon mass,
$\Delta=2$, produces the best results.

We finally have included the $\pi NN$ and $\rho NN$ coupling
constants.  Upon calculating the coupling constants, it is essential
to use the same $z_m$ for mesons and nucleons, so that whole
information about the wavefunctions are not lost in the course of the
calculation. We have performed the global fitting procedure in which
we obtained the best fit with the values of the parameters:
$z_m=(285\,\mathrm{MeV})^{-1}$, $\sigma=(227\,\mathrm{MeV})^3$, and
$g=-9.6$. The mass spectra of the nucleon and the $\rho$ meson are in
relatively good agreement with the experimental data within
$10-20\,\%$, whereas the $\pi NN$ and $\rho NN$ coupling constants
underestimated by about $50\,\%$.

In order to improve the present results, one might consider higher
dimensional operators~\cite{KKW}, or a finite UV cutoff~\cite{ET}.

\acknowledgments

H.Ch.K. and U.Y. are grateful for the warm hospitality during their
visit to APCTP, where a part of the work has been done. The work of
U.~Yakhshiev is partially supported by AvH foundation. The present
work is also supported by Basic Science Research Program through the
National Research Foundation of Korea (NRF) funded by the Ministry
of Education, Science and Technology (grant number: 2009-0073101).
Y.K. acknowledges the Max Planck Society(MPG) and the Korea Ministry
of Education, Science and Technology(MEST) for the support of the
Independent Junior Research Group at the Asia Pacific Center for
Theoretical Physics (APCTP).


\begin{thebibliography}{999}

\bibitem{Maldacena:1997re}
  J.~M.~Maldacena,
{\em The large N limit of superconformal field theories and supergravity,}
  \atmp{2}{1998}{231}
  [\ijtp{\bf 38}{1999}{1113}]  [\hepth{9711200}].

\bibitem{Gubser:1998bc}
  S.~S.~Gubser, I.~R.~Klebanov and A.~M.~Polyakov,
{\em Gauge theory correlators from non-critical string theory,}
  \plb{428}{1998}{105}  [\hepth{9802109}].

\bibitem{Witten:1998qj}
  E.~Witten,
{\em Anti-de Sitter space and holography,}
  \atmp{2}{1998}{253}  [\hepth{9802150}].

\bibitem{Erdmenger:2007cm}
  J.~Erdmenger, N.~Evans, I.~Kirsch and E.~Threlfall,
{\em Mesons in Gauge/Gravity Duals - A Review,}
  \newjournal{Eur.\ Phys.\ J. \bf A}{epja}{35}{2008}{81}
  [\arXivid{0711.4467}].

\bibitem{Karch:2002sh}
  A.~Karch and E.~Katz,
{\em Adding flavor to AdS/CFT},
{\em JHEP} {\bf 0206} (2002) 043
  [\hepth{0205236}].

\bibitem{Kruczenski:2003be}
  M.~Kruczenski, D.~Mateos, R.~C.~Myers and D.~J.~Winters,
{\em Meson spectroscopy in AdS/CFT with flavour},
{\em JHEP} {\bf 0307} (2003) 049 [\hepth{0304032}].

\bibitem{Kruczenski:2003uq}
  M.~Kruczenski, D.~Mateos, R.~C.~Myers and D.~J.~Winters,
{\em Towards a holographic dual of large-N(c) QCD},
{\em JHEP} {\bf 0405} (2004) 041
  [\hepth{0311270}].

\bibitem{Sakai:2004cn}
  T.~Sakai and S.~Sugimoto,
{\em Low energy hadron physics in holographic QCD},
{\em  Prog.\ Theor.\ Phys.}  {\bf 113} (2005) 843
  [\hepth{0412141}].

\bibitem{Sakai:2005yt}
  T.~Sakai and S.~Sugimoto,
{\em More on a holographic dual of QCD},
{\em  Prog.\ Theor.\ Phys.} {\bf 114} (2005) 1083
  [\hepth{0507073}].

\bibitem{Erlich:2005qh}
  J.~Erlich, E.~Katz, D.~T.~Son and M.~A.~Stephanov,
  {\em QCD and a holographic model of hadrons},
{\em Phys.\ Rev.\ Lett.} {\bf 95} (2005) 261602
  [\hepph{0501128}].

\bibitem{DaRold:2005zs}
  L.~Da Rold and A.~Pomarol,
{\em Chiral symmetry breaking from five dimensional spaces},
{\em  Nucl.\ Phys.} {\bf B 721} (2005) 79
  [\hepph{0501218}].

\bibitem{DaRold:2005vr}
  L.~Da Rold and A.~Pomarol,
{\em The scalar and pseudoscalar sector in a five-dimensional approach to
chiral symmetry breaking},
{\em JHEP} {\bf 0601} (2006) 157
  [\hepph{0510268}].

\bibitem{Batell:2008zm}
  B.~Batell and T.~Gherghetta,
{\em Dynamical Soft-Wall AdS/QCD,}
  \prd{78}{2008}{026002} [\arXivid{0801.4383}].
\bibitem{Batell:2008me}
  B.~Batell, T.~Gherghetta and D.~Sword,
{\em The Soft-Wall Standard Model,}
  \prd{78}{2008}{116011}  [\arXivid{0808.3977}].
\bibitem{Andreev:2006vy}
  O.~Andreev,
{\em  1/q**2 corrections and gauge / string duality,}
  \prd{73}{2006}{107901}  [\hepth{0603170}].
\bibitem{Contino:2004vy}
  R.~Contino and A.~Pomarol,
{\em Holography for fermions,}
  \jhep{0411}{2004}{058} [\hepth{0406257}].
\bibitem{Gherghetta:2000qt}
  T.~Gherghetta and A.~Pomarol,
{\em Bulk fields and supersymmetry in a slice of AdS,}
  \npb{586}{2000}{141}  [\hepph{0003129}].
\bibitem{Mueck:1998iz}
  W.~M\"uck and K.~S.~Viswanathan,
{\em Conformal field theory correlators from classical field theory on  anti-de
  Sitter space. II: Vector and spinor fields,}
  \prd{58}{1998}{106006}  [\hepth{9805145}].
\bibitem{Hong:2006ta}
  D.~K.~Hong, T.~Inami and H.~U.~Yee,
{\em  Baryons in AdS/QCD,}
  \plb{646}{2007}{165} [\hepph{0609270}].
\bibitem{Arutyunov:1998ve}
  G.~E.~Arutyunov and S.~A.~Frolov,
{\em On the origin of supergravity boundary terms in the AdS/CFT
  correspondence,}
  \npb{544}{1999}{576}  [\hepth{9806216}].
\bibitem{Henningson:1998cd}
    M.~Henningson and K.~Sfetsos,
{\em  Spinors and the AdS/CFT correspondence,}
  \plb{431}{1998}{63}  [\hepth{9803251}].
\bibitem{Maru:2009ux}
  N.~Maru and M.~Tachibana,
  {\em Meson-Nucleon Coupling from AdS/QCD,}
  \arXivid{0904.3816}.
\bibitem{Kim:2008qh}
  H.-Ch.~Kim and Y.~Kim, \textit{Hybrid exotic meson with $J^{PC}=1^{-+}$
    in AdS/QCD},
\jhep{0901}{2009}{034} [\arXivid{0811.0645}].
\bibitem{Hong:2007tf}
  D.~K.~Hong, H.-Ch.~Kim, S.~Siwach and H.~U.~Yee,
\textit{The Electric Dipole Moment of the Nucleons in Holographic QCD},
\jhep{0711}{2007}{036} [\arXivid{0709.0314}].
\bibitem{Kim:2007xi}
  Y.~Kim, C.~H.~Lee and H.~U.~Yee,
\textit{Holographic Nuclear Matter in AdS/QCD},
\prd{77}{2008}{085030} [\arXivid{0707.2637}].

\bibitem{KP2009}
M.H. Park, \textit{Perturbative Approach to Baryons in AdS/QCD}, Ms. D. thesis, Feburary 2009;
 N. Kim and M.H. Park, \textit{The Spectrum Of Baryons In AdS/QCD}, talk give at
 \textit{Korean Physical Society meeting, 23-24 April, 2009, Daejeon, Korea}.

\bibitem{Nambu:1961tp}
  Y.~Nambu and G.~Jona-Lasinio,
  \textit{Dynamical model of elementary particles based on an analogy with
superconductivity. I},
\pr{122}{1961}{345}.
\bibitem{Klevansky:1992qe}
  S.~P.~Klevansky, \textit{The Nambu-Jona-Lasinio model of quantum
    chromodynamics},
\rmp{64}{1992}{649}.
\bibitem{Christov:1995vm}
  C.~V.~Christov {\it et al.}, \textit{Baryons as non-topological
    chiral solitons}, \textit{Prog.\ Part.\ Nucl.\ Phys.} {\bf 37}
  (1996) 91 [\hepph{9604441}].

\bibitem{PMWHW}
M. Procura, B.U. Musch, T. Wollenweber, T.R. Hemmert, and W. Weise,
 Phys. Rev. {\bf D73}, 114510 (2006)[hep-lat/0603001 ].


\bibitem{KKW} H. R. Grigoryan,  Phys. Lett. {\bf B662}, 158 (2008), arXiv:0709.0939 [hep-ph];
Y. Kim, P. Ko, and X.-H. Wu, JHEP {\bf 0806}, 094 (2008), arXiv:0804.2710 [hep-ph].

\bibitem{ET} N. Evans and A. Tedder, Phys. Lett. {\bf B642}, 546 (2006), e-Print: hep-ph/0609112.

\end{thebibliography}
\end{document}